\documentclass[acmsmall,screen]{acmart}

\AtBeginDocument{%
  }
    
\usepackage{threeparttable}
\usepackage{pifont}
\usepackage{float}
\usepackage{subfigure}

\setcopyright{acmlicensed}
\copyrightyear{2026}
\acmYear{2026}
\acmDOI{XXXXXXX.XXXXXXX}

\acmJournal{JACM}
\acmVolume{37}
\acmNumber{4}
\acmArticle{111}
\acmMonth{10}

\usepackage{multirow}
\usepackage{array}
\usepackage{tabularx}
\usepackage{placeins}
\usepackage{needspace}
\usepackage{enumitem}
\usepackage{pifont}
\usepackage{longtable}
\usepackage{lscape}
\usepackage{rotating}
\usepackage{adjustbox}
\usepackage[most]{tcolorbox}
\usepackage{fvextra}
\usepackage{tikz}
\usetikzlibrary{arrows.meta,positioning,shapes.geometric}

\begin{document}

\title{\benchmark{}: Benchmarking Multilingual Code Generation from Functions to Repositories}

\author{Bowen Yang}
\affiliation{\institution{Tianjin University}\city{Tianjin}\country{China}}
\email{3023244317@tju.edu.cn}

\author{Jiajun Jiang}
\authornote{Corresponding author.}
\affiliation{\institution{Tianjin University}\city{Tianjin}\country{China}}
\email{jiangjiajun@tju.edu.cn}

\author{Luxue Yu}
\affiliation{\institution{Tianjin University}\city{Tianjin}\country{China}}
\email{ylx303@tju.edu.cn}

\author{Yihao Wang}
\affiliation{\institution{Tianjin University}\city{Tianjin}\country{China}}
\email{wyh132465@tju.edu.cn}

\author{Fengjie Li}
\affiliation{\institution{Tianjin University}\city{Tianjin}\country{China}}
\email{fengjie@tju.edu.cn}

\author{Dong Wang}
\affiliation{\institution{Tianjin University}\city{Tianjin}\country{China}}
\email{dong_w@tju.edu.cn}

\renewcommand{\shortauthors}{Yang et al.}

\newcolumntype{L}[1]{>{\raggedright\arraybackslash}p{#1}}
\newcolumntype{C}[1]{>{\centering\arraybackslash}p{#1}}
\newcolumntype{Y}{>{\raggedright\arraybackslash}X}

\newcommand{\polytablefont}{\footnotesize}
\newcommand{\polytablespacing}{\setlength{\tabcolsep}{5.2pt}\renewcommand{\arraystretch}{1.08}}

\newcommand{\checkmarkicon}{\ding{51}}
\newcommand{\crossmarkicon}{\ding{55}}

\newcommand{\benchmark}[1]{\textsc{PolyCodeEval}}

\newcommand{\homepage}[1]{\url{https://polycodeeval.github.io/PolyCodeEval}}

\newcommand{\polyinsight}[1]{%
  \vspace{4pt}%
  \noindent\fcolorbox{black}{gray!18}{%
    \parbox{\dimexpr\columnwidth-2\fboxsep-2\fboxrule\relax}{\footnotesize #1}%
  }%
  \vspace{4pt}%
}

\newtcolorbox[auto counter]{summary}[1][]{%
    colback=gray!10,
    colframe=gray!80,
    leftrule=3pt,
    rightrule=0pt,
    toprule=0pt,
    bottomrule=0pt,
    arc=0pt,
    boxsep=0pt,
    left=8pt,
    right=8pt,
    top=4pt,
    bottom=4pt,
    title={},
    fonttitle=\bfseries,
    coltitle=black,
    attach title to upper={\ },
    #1
}

\newcommand{\distance}{3pt}
\setlength{\textfloatsep}{\distance}
\setlength{\floatsep}{\distance}
\setlength{\dbltextfloatsep}{\distance}
\setlength{\dblfloatsep}{\distance}

\begin{abstract}
As large language models increasingly move toward repository-level software engineering, existing code-generation benchmarks remain fragmented across language coverage, task granularity, and evaluation protocols, impeding systematic comparison. 
To address this gap, we present \benchmark{}, a unified multilingual and multi-granularity benchmark for code generation. It comprises 2,590 code generation tasks spanning functions to repositories, derived from 58 real, executable open-source repositories in five programming languages. All tasks are evaluated under a unified execution-based protocol with integration procedures tailored to their generation targets.
Building on this benchmark, we evaluate frontier large language models, state-of-the-art specialized methods, and general coding agents. Our results show that existing approaches still struggle to correctly generate complete code fragments across granularities and languages. Specifically, the studied methods generate at most 71.7\%, 76.7\%, and 31.0\% correct functions, files, and repositories, respectively, with performance varying widely across languages. Paired experiments further show that implementation context from related functions in the same file improves the executable correctness of function generation. Method rankings also vary across task granularities and programming languages, highlighting the importance of multilingual, multi-granularity evaluation for comprehensively assessing code generation capabilities.

\end{abstract}




\maketitle

\section{Introduction}

Benchmarks play a central role in code generation research because they determine which capabilities are measured, which forms of progress can be observed, and which system behaviors are ultimately optimized~\cite{joel2024survey, dong2025survey, jiang2026survey}. 
As code generation moves from local completion toward repository-grounded software engineering tasks, evaluation must also move beyond isolated snippets and textual similarity. The key question is now whether a model can produce cross-file, internally consistent, executable software artifacts under realistic repository constraints.

Existing benchmarks have substantially advanced code-generation evaluation, but they remain fragmented in language coverage, task granularity, and validation protocol. Early benchmarks such as HumanEval~\cite{ref_humaneval} and MBPP~\cite{ref_mbpp} focus on short functional programs or algorithmic snippets. 
Recent benchmarks extend evaluation toward repository-level completion, issue resolution, and software engineering agents~\cite{ref_repocoder,ref_swebench,ref_chatdev}. However, many benchmarks still concentrate on a single language, especially Python, and multilingual settings often remain centered on function- or statement-level completion~\cite{ref_multipl_e,ref_crosscodeeval}. Different granularities are also commonly built from different datasets~\cite{ref_swebench,ref_projecteval,ref_projdevbench}, which makes it difficult to compare how the same underlying repository behaves when the target changes from whole-repository generation to file- or function-level completion.

This limitation becomes increasingly important because code-generation systems now operate under substantially different engineering conditions.
The generation settings range from producing a complete runnable repository from scratch, to filling a missing file in an existing repository, to implementing only a target function. These settings differ in visible context, integration constraints, and dominant failure modes.
However, there still lacks a unified benchmark spanning these levels for studying how model behavior changes as the generation target becomes finer-grained while repository consistency requirements remain strict.

To address this gap, this paper introduces \benchmark{}, a multilingual code generation benchmark and execution-based evaluation platform constructed from complete, executable Oracle repositories.
\benchmark{} uses 58 real open-source repositories in five programming languages, including Python, JavaScript, Java, Go, and C++, as a shared data source, and defines four related task levels, naming from L0 to L3, according to the engineering information visible to the model and the required generation target. Specifically, L0 provides only software requirements for generating a complete repository from an empty workspace. L1 retains the same complete-repository generation objective but additionally provides a repository skeleton that preserves the directory structure, expected files, public interfaces, and necessary build and configuration files while removing concrete implementations. L2 retains the repository environment outside the target file and asks the model to generate one complete target file. L3 focuses on generating an individual function based on the given signature and its functional description.
The resulting benchmark contains 58 L0 tasks, 58 L1 tasks, 150 L2 tasks, and 2,324 L3 tasks, for a total of 2,590 tasks. Because all four levels are derived from the same Oracle repositories, \benchmark{} enables controlled cross-level comparison over a shared repository population.

\benchmark{} uses a unified parsing and execution pipeline to construct repository-, file-, and function-level tasks from executable source repositories. To support reliability and reproducibility, we validate L0/L1 tasks with interface-level functional tests, evaluate L2/L3 tasks by backfilling generated content into the corresponding Oracle-derived repository structure and then running build and test execution, audit and augment test coverage across all source repositories, and verify every task path through Oracle execution before model evaluation. These procedures ensure task executability, evaluation consistency, and reliable analysis of generation failures.

On top of this platform, we conduct a large-scale empirical study of code generation methods across task granularities, programming languages, and repository constraints. At L3, we compare LLM-direct generation, the state-of-the-art RepoCoder~\cite{ref_repocoder}, HCP~\cite{ref_hcp}, and AlignCoder~\cite{ref_aligncoder}, using GPT-5.4, Claude Sonnet 4.6, and DeepSeek-V4 Pro as generator backbones. At L2, we compare direct generation with GPT-5.4 and Claude Sonnet 4.6 against agent-based generation with Codex and Claude Code. L1 evaluates Codex and Claude Code on complete-repository generation with a provided repository skeleton. L0 further includes a multilingual adaptation of the latest ChatDev~\cite{ref_chatdev} alongside Codex and Claude Code for repository-level comparison. These experiments allow \benchmark{} to compare direct generation, repository-aware methods, and coding agents under a unified execution-based protocol, and to examine how their performance and failure modes vary across task levels.

\textbf{The results reveal level-specific capability bottlenecks}. \ding{182}At the repository level (L0/L1), build-successful repositories typically pass a high proportion of tests, while the overall full-pass rate remains low, indicating that complete-repository generation must prioritize build configuration, dependency management, interface consistency, and engineering integration. Providing an additional repository skeleton does not yield a consistent overall improvement and can adversely affect some metrics. \ding{183}At the file level (L2), the main gains from Codex arise from improved build success, further showing that accurate modeling of repository interfaces, dependency relations, and existing implementations is central to file-level generation. \ding{184}At the function level (L3), RepoCoder with GPT-5.4 achieves a 71.7\% full-pass rate, compared with 48.8\% for LLM-direct generation, while more elaborate context-processing methods do not exhibit consistent advantages. \ding{185}The paired comparison further shows that implementation context from related functions in the same file improves the executable correctness of the target function. These results associate performance gaps at different granularities with concrete directions for system improvement: repository-level generation requires stronger engineering integration, file-level generation requires better understanding of interfaces and dependencies, and function-level generation requires more effective context selection and organization. 
These findings motivate future work to focus on context utility and alignment between generated artifacts and real engineering environments.

In summary, the major contributions of this paper are threefold:
\begin{enumerate}
\item \textbf{Benchmark.} We present \benchmark{}, a unified multilingual and multi-granularity benchmark for code generation, spanning complete-repository generation, skeleton-conditioned repository generation, file-level generation, and function-level generation.
\item \textbf{Unified Evaluation Platform.} We construct all tasks from the same set of executable Oracle repositories and evaluate them under a unified execution-based protocol with level-appropriate integration paths, supported by coverage auditing, Oracle task-path verification, and shared evaluation infrastructure.
\item \textbf{Comprehensive Empirical Study.} We provide a comprehensive empirical evaluation of direct generation, repository-aware methods, and agent-based methods on this platform, and report a set of findings on granularity effects, cross-language variation, build-related bottlenecks, and repository-context usage that can guide future research.
\end{enumerate}

\section{Related Work and Limitations}

\begin{table*}[!t]
\caption{Representative code generation and software engineering benchmarks as of September 2026}
\label{tab:benchmark-compare}
\centering
\polytablefont
\setlength{\tabcolsep}{2.8pt}
\renewcommand{\arraystretch}{1.08}
\begin{tabularx}
{\textwidth}{@{}L{2.4cm}Y C{1.2cm}C{1.0cm}C{1.0cm}C{1.0cm}C{0.72cm}C{0.85cm}@{}}
\toprule
\multirow{2}{*}{Benchmark} & \multirow{2}{*}{Task description} & \multicolumn{4}{c}{Code Generation Levels (Granularity)} & \multirow{2}{*}{Lang.} &
\multirow{2}{*}{Eval.} \\
\cmidrule(lr){3-6}
& & \shortstack{L0: Repo.} & \shortstack{L1: Skel.} & \shortstack{L2: File} & \shortstack{L3: Func.} &  & \\
\midrule
HumanEval \cite{ref_humaneval} & Function completion & -- & -- & -- & 164 & 1 & Exec. \\
MBPP \cite{ref_mbpp} & Function implementation & -- & -- & -- & 974 {\scriptsize(7)} & 1 & Exec. \\
DS-1000 \cite{ref_ds1000} & Data-science code generation & -- & -- & -- & 1000 {\scriptsize(5)} & 1 & Exec. \\
BigCodeBench \cite{ref_bigcodebench} & Function-call generation & -- & -- & -- & 1140 {\scriptsize(10)} & 1 & Exec. \\
MultiPL-E \cite{ref_multipl_e} & Multilingual functions & -- & -- & -- & 10678 & 18 & Exec. \\
RepoEval \cite{ref_repocoder} & Repository function completion & -- & -- & -- & 373 & 1 & Exec. \\
CrossCodeEval \cite{ref_crosscodeeval} & Cross-file statement completion & -- & -- & -- & 9928 & 4 & Text. \\
xCodeEval \cite{ref_xcodeeval} & Multilingual evaluation & -- & -- & -- & 11,638 & 11 & Exec. \\
DevEval \cite{ref_deveval} & Repository function generation & -- & -- & -- & 1825 & 1 & Exec. \\ 
HumanEvo~\cite{zheng2025humanevo} & Evolution-aware function generation & -- & -- & -- & 400 {\scriptsize(24)} & 2 & Exec.\\
M2RC-EVAL \cite{ref_m2rc} & Multilingual statement completion & -- & -- & -- & 10800 & 18 & Text. \\
SWE-bench \cite{ref_swebench} & Repository issue fixing & -- & -- & 2294 {\scriptsize(33)} & -- & 1 & Exec. \\
ProjectEval \cite{ref_projecteval} & Repository + skeleton completion & 20 {\scriptsize(402)} & 20 {\scriptsize(402)} & -- & -- & 1 & Exec. \\
ProjDevBench \cite{ref_projdevbench} & Repository-level development & 20 & -- & -- & -- & 2 & Exec.  \\
\midrule
\textbf{PolyCodeEval} & \textbf{Multi-granularity generation} & {\scriptsize\textbf{58 (2464)}} & {\scriptsize\textbf{58 (2464)}} & {\scriptsize\textbf{150 (898)}} & {\scriptsize\textbf{2324 (29)}} &
\textbf{5} & \textbf{Exec.} \\
\bottomrule
\end{tabularx}

\begin{minipage}{\textwidth}
\footnotesize
\textit{L0–L3:}  Number of tasks at each level; values in parentheses denote average LoC;
\textit{Lang.:} number of programming languages.\\
\textit{Eval.:} Exec. = execution-based test validation; Text. = non-executed textual-similarity evaluation.
\end{minipage}
\end{table*}

This section reviews related work and identifies limitations in existing studies. LLMs have shown promising code generation results, prompting many evaluation benchmarks. To clarify current progress, Table~\ref{tab:benchmark-compare} summarizes representative benchmarks, mainly from top-tier 2023–2026 venues, including each benchmark’s name, tasks, code generation granularity, supported languages, and evaluation method.

\textbf{Function-Level Benchmarks}. Early code-generation benchmarks focused on isolated algorithmic functions. HumanEval~\cite{ref_humaneval} introduced 164 hand-crafted Python problems with function signatures, docstrings, and unit tests (averaging 7.7 cases), becoming a standard. MBPP~\cite{ref_mbpp} expanded this with 974 crowd-sourced entry-level Python problems covering fundamentals and standard libraries, each with a task description, solution, and three tests. DS-1000~\cite{ref_ds1000} further extended to data science domain, compising 1,000 problems across seven Python libraries such as NumPy, Pandas, and SciPy. While these benchmarks are influential, they test isolated function generation rather than realistic software engineering, and their evaluation is predominantly confined to Python.

\textbf{Function/File-Level Benchmarks with Repository Contexts}. 
To bridge toward real-world software development, function-level benchmarks with repository contexts have emerged. RepoEval~\cite{ref_repocoder} targets Python next-line completion across line, API invocation, and function-body granularity using 14 repositories. CrossCodeEval~\cite{ref_crosscodeeval} addresses the limitation of single-file tasks by introducing a multilingual benchmark for cross-file code completion in Python, Java, TypeScript, and C\#, requiring cross-file context via static analysis. DevEval~\cite{ref_deveval} provides 1,825 manually annotated samples from 115 repositories across 10 domains, including requirements, reference code, and dependencies. SWE-bench~\cite{ref_swebench} advances to real GitHub issues in Python, requiring agents to navigate codebases and generate patches that pass existing tests, and has expanded into a task family including SWE-bench Multilingual, covering 9 languages beyond Python.

\textbf{Repository-Level Benchmarks}. 
More recently, benchmarks have moved to repository-level generation. ProjectEval~\cite{ref_projecteval}, LLM-built with human review, varies inputs (natural language or code skeletons) and evaluates generated repositories via simulated user interaction and code similarity; it finds that systematic engineering, whole-repository understanding, and comprehensive analysis are key for practical LLM agents. ProjDevBench~\cite{ref_projdevbench} is end-to-end, giving agents software requirements and evaluating resulting repositories through Online Judge (OJ) testing and LLM-assisted code review across architecture design, functional correctness, and iterative refinement. With 20 problems in 8 categories, it reports 27.38\% acceptance, showing agents handle basic functionality but struggle with complex system design, time-complexity optimization, and resource management.

\textbf{Multilingual Benchmarks}. 
To address language coverage, MultiPL-E~\cite{ref_multipl_e} translates unit test-driven code-generation benchmarks into multiple languages, extending HumanEval and MBPP to 18 languages across diverse paradigms. xCodeEval~\cite{ref_xcodeeval}, the largest executable multilingual multitask benchmark to date, contains 25M document-level examples from about 7.5K problems in up to 11 languages, with 7 tasks covering understanding, generation, translation, and retrieval, evaluated by a multilingual execution engine. Despite these advances, benchmarks jointly covering many languages and scopes from functions to files and complete repositories remain limited.

\textit{Limitations of Existing Benchmarks}. Collectively, these benchmarks provide an important foundation for evaluating code generation systems but exhibit several limitations:

\begin{enumerate}
    \item \textbf{Uneven language coverage and task forms}. 
    Widely used benchmarks such as HumanEval, MBPP, DevEval, and SWE-bench remain centered on Python or a single language. Although MultiPL-E and xCodeEval broaden language coverage, they mainly evaluate local code units such as functions or statements. Few benchmarks jointly cover multiple languages and scopes from functions to files and complete repositories.
    \item \textbf{Lack of controlled cross-granularity organization}.
    Existing benchmarks usually target one granularity, with function-, file-, and repository-level tasks drawn from different repositories and defined by different input formats, construction procedures, and evaluation protocols. Consequently, performance differences are thus confounded by data source, language, domain, and granularity, making it hard to isolate generation scope or analyze capability boundaries across granularities.

    \item \textbf{Inconsistent assessment of task completion}. 
    Some benchmarks rely on exact match or textual similarity without building or executing generated code. Such metrics offer limited evidence on dependency completeness, interface compatibility, build success, and cross-file integration, while penalizing functionally equivalent alternatives. For file- and repository-level tasks, build outcomes, execution behavior, and functional tests are more direct evidence of requirement satisfaction.
\end{enumerate}

\textit{These limitations give rise to three interrelated benchmark-construction challenges.} \textbf{First}, cross-granularity evaluation requires comparable tasks derived from a common repository population, while controlling potential confounding factors such as language, application domain, and functional objective so that performance differences can be more clearly associated with generation scope and input constraints. \textbf{Second}, multilingual evaluation must accommodate substantial variation in repository layout, dependency management, build systems, testing frameworks, and artifact-integration procedures while preserving consistent task definitions and metric semantics. \textbf{Third}, execution-based evaluation requires reliable validation paths for tasks at different granularities, together with adequate tests, coverage inspection, and Oracle validation to identify both functional defects and engineering-integration failures.

In this paper, we propose \benchmark{}, which addresses these challenges by constructing code generation tasks from individual functions to complete repositories spanning five widely-used programming languages. Tasks at different levels share a common repository and functional foundation while varying generation scope, visible structure, and contextual information in a controlled manner. All generated artifacts are integrated into complete repository environments and evaluated through build and functional testing, enabling systematic comparison of code generation capabilities across languages and granularities.

\section{\benchmark{} Benchmark}

\subsection{Benchmark Overview}
\label{sec:overview}
\benchmark{} is a \textit{multilingual}, \textit{multi-granularity} benchmark for evaluating repository-grounded code generation under realistic engineering constraints. It is constructed from 58 executable open-source repositories in Python, JavaScript (JS), Java, Go, and C++, and covers four generation levels from complete repositories to individual functions.
A key design principle of \benchmark{} is that all task levels are derived from the same Oracle repository pool, enabling more controlled comparison across generation granularities while reducing variation caused by unrelated data sources. The benchmark further adopts a unified execution-based evaluation framework, in which generated artifacts are integrated into realistic repository environments and assessed through \textit{build} and \textit{test execution} rather than textual similarity.
The final benchmark contains \textbf{2,590 tasks} across repository-, file-, and function-level generation.

\subsection{Multi-Granularity Task Definition}
\begin{figure}[!t]
\centering
\includegraphics[width=0.95\textwidth]{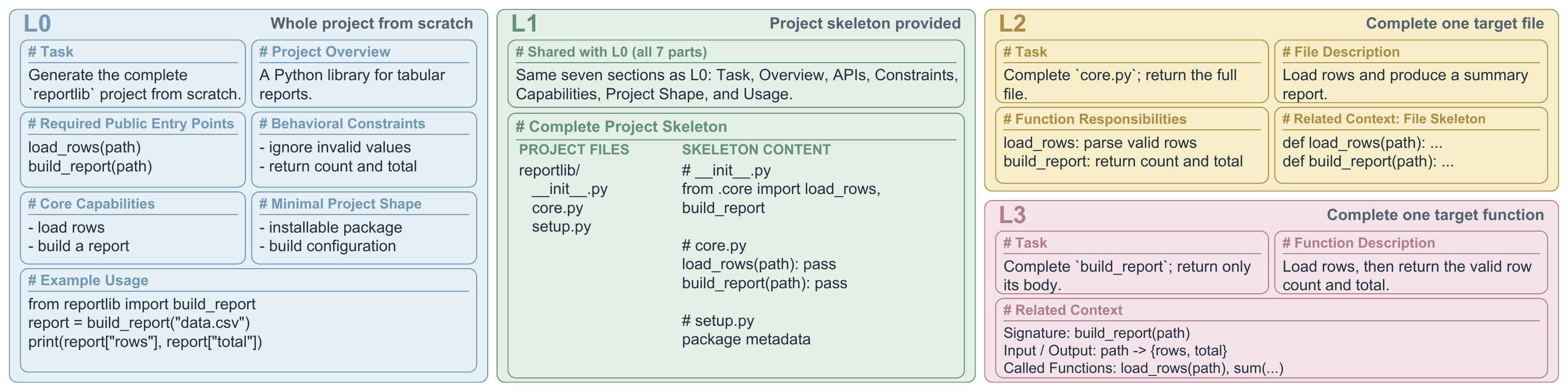}
\caption{Task definition and input structure of examples for L0--L3.}
\label{fig:prompt-structure-examples}
\end{figure}

\benchmark{} organizes code generation into four task levels, denoted L0–L3, according to two dimensions: the scope of the artifact to be generated and the amount of repository information available to the model. 
All four levels are derived from executable Oracle repositories, but they expose different forms of structural and implementation context. 
This design covers a progression from complete-repository generation to localized function implementation while maintaining a common repository-grounded setting. Figure~\ref{fig:prompt-structure-examples} presents the example for each task level.

\begin{itemize}[leftmargin=10pt]
\item  \textbf{L0: Repository Generation.}
It represents complete-repository generation from a natural-language software specification. 
The model starts from an empty workspace without access to the original repository structure or implementation.
The task is to construct a complete, executable software repository that satisfies the specified functional requirements. 
This includes not only source-code implementation but also the engineering artifacts required for execution, such as directory organization, dependency declarations, and build or configuration files.
L0 therefore captures the most open-ended setting in \benchmark{}, requiring the model to make both implementation and repository-organization decisions from high-level requirements.

\item  \textbf{L1: Skeleton-Conditioned Repository Generation.}
L1 retains the same complete-repository generation objective as L0 but additionally exposes the structural skeleton of the target repository.
The provided skeleton preserves information such as the directory hierarchy, expected source files, public interfaces, type declarations, and necessary engineering or configuration structures, while concrete implementation logic is removed. The model must complete the missing implementation while respecting the predefined repository organization and interfaces.

\item \textbf{L2: File-Level Generation.}
L2 considers code generation at the file level within an existing repository.
The surrounding repository environment and the code structure in the file is preserved, while the detailed implementation of a selected target file is removed. The model is required to generate the complete target file while remaining compatible with the existing codebase.

\item  \textbf{L3: Function-Level Generation.}
L3 targets individual-function generation. Specifically, the target implementation of the function is removed while its repository context and interface information are retained, and the model must produce an implementation that satisfies the intended behavior and remains compatible with surrounding code. Though L3 is \benchmark{}'s smallest generation unit, it differs from isolated algorithmic benchmarks because the function remains embedded in a real repository.

\end{itemize}

\subsection{Benchmark Construction}
\label{sec:collection}

\textbf{Repository Sources.}
We collect candidate repositories from two complementary sources: existing code-generation benchmarks and GitHub. Specifically, we retain 18 source repositories from DevEval~\cite{ref_deveval} and CodeProjectEval~\cite{ref_projectgen}, and additionally collect more than 100 candidate repositories from GitHub, yielding an initial pool of over 120 repositories. This broadens programming-language and repository-scale coverage, particularly for languages insufficiently represented by the existing benchmarks. The GitHub candidates primarily use one of the five target languages, generally have more than 500 stars, and span different LoC ranges. To balance the repository distribution across languages and complexity levels, we also include a small number (i.e., five) of repositories that pass our manual review and satisfy the same downstream admission criteria.
For repositories originating from prior benchmarks, we reuse only the source repositories rather than their original task definitions or prompts; all \benchmark{} tasks are reconstructed through our unified pipeline.

\begin{figure}[t]
    \centering
    \includegraphics[width=1\textwidth]{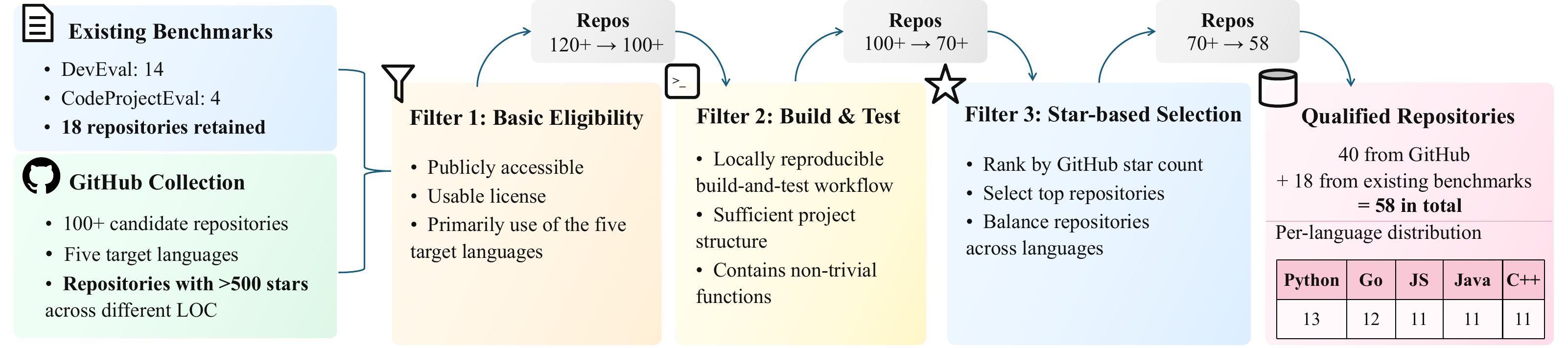}
    \caption{\benchmark{}'s repository collection and qualification pipeline.}
    \label{fig:data-collection-pipeline}
\end{figure}

\smallskip
\noindent
\textbf{Repository Qualification.}
All candidates are screened using the same admission criteria. After basic eligibility screening for public accessibility, usable licenses, and primary use of one of the five target languages, more than 100 repositories remain. Further validation of locally reproducible build-and-test workflows, sufficient repository structure, source files, and non-trivial functions reduces the pool to more than 70 candidates suitable for multi-granularity task construction.
We exclude repositories whose execution depends on unavailable private services, inaccessible credentials, or mandatory manual interaction.
Qualified candidates are further normalized for reproducible execution. 
We separate production code, tests, fixtures, and runtime resources; exclude vendored dependencies, package caches, generated artifacts, and unrelated test resources; and normalize dependency declarations, build configurations, resource paths, and evaluation layouts. 
The normalized original implementation is then treated as the Oracle repository and retained only if it builds successfully, passes its native test suite, supports coverage auditing, and provides the artifacts required for task construction. Figure~\ref{fig:data-collection-pipeline} presents our repository filtering process in detail.

After this process, we obtain 58 qualified Oracle repositories, including 18 from existing benchmarks and 40 from GitHub, covering 13 Python, 12 Go, 11 JavaScript, 11 Java, and 11 C++ repositories.

\noindent
\textbf{Multi-Granularity Task Extraction.} 
We next derive the four task levels defined in Section 3.2 from the qualified Oracle repositories. The key principle is to preserve the same underlying software systems while varying the generation scope and the information available to the model.

\begin{itemize}[leftmargin=10pt]
\item \textbf{For L0 and L1}, each qualified Oracle repository forms one repository-level generation task. L0 provides only the software requirements and requires generation of the complete repository from an empty workspace. L1 uses the same repository-level target but additionally provides the full repository skeleton. Because both levels are derived from the same 58 repositories, they each contain \textbf{58 repository-level tasks}, with an average target size of 2,464.3 LOC.
\item \textbf{For L2}, we identify target source files with sufficient local complexity. Each selected file must contain at least three non-trivial functions. 
The implementation of the target file is removed while the file structure and the rest of the repository is preserved for context construction and execution. This process yields \textbf{150 file-level tasks}. Each task requires generation of a complete file from its file skeleton and repository context, with an average target size of 898.1 LOC.
\item \textbf{For L3}, we traverse non-trivial functions in the Oracle repositories and construct one task for each eligible function. A non-trivial function is defined as having a body of at least three lines of code and not being a simple getter- or setter-style accessor. The target function body is removed, while its signature, source location, and metadata required for reintegration are retained. 
This produces \textbf{2,324 function-level tasks}, with an average target size of 28.8 LOC.
\end{itemize}

Overall, \benchmark{} contains 2,590 tasks spanning complete-repository, skeleton-conditioned repository, file-level, and function-level generation. 
Table~\ref{tab:levels} presents the task summary and task distribution of different complexities across languages according to targeted LOC.

\subsection{Input Construction and Quality Assurance}
\label{sec:improvement}

Task extraction determines the artifact to be generated, but the retained repository information must still be transformed into standardized inputs that can be consumed by evaluated models. Because repository-level and local-completion tasks expose substantially different forms of information, \benchmark{} adopts two input-construction pipelines: one for L0/L1 and another for L2/L3.

\begin{table*}[tbp]
\caption{\benchmark{} task definitions and dataset composition.}
\label{tab:levels}
\centering
\scriptsize
\setlength{\tabcolsep}{3.2pt}
\renewcommand{\arraystretch}{0.96}
\begin{threeparttable}
\begin{tabular*}{\textwidth}{@{\extracolsep{\fill}}llllrrc@{}}
\toprule
\multicolumn{6}{l}{\textit{(a) Task definitions}} \\
\midrule
Level & Task description & Input & Output & Tasks & Avg. LOC \\
\midrule
L0 & Repository generation & Project Requirement Doc(PRD) only & Complete repository & 58 & 2,464.3 \\
L1 & Repository-level completion & PRD + full repository skeleton & Complete repository & 58 & 2,464.3 \\
L2 & File-level completion & File skeleton + context & Complete file & 150 & 898.1 \\
L3 & Function-level completion & Function description + context & Complete function & 2,324 & 28.8 \\
\bottomrule
\end{tabular*}

\setlength{\tabcolsep}{2.7pt}
\begin{tabular*}{\textwidth}{@{\extracolsep{\fill}}l*{15}{c}@{}}
\toprule
\multicolumn{16}{l}{\textit{(b) Tasks by language and target size}} \\
\midrule
& \multicolumn{5}{c}{L0/L1: repository} & \multicolumn{5}{c}{L2: file} & \multicolumn{5}{c}{L3: function} \\
\cmidrule(lr){2-6}\cmidrule(lr){7-11}\cmidrule(l){12-16}
Language & S & M & L & XL & Total & S & M & L & XL & Total & S & M & L & XL & Total \\
\midrule
Python     & 8 & 2 & 3 & 0 & 13 & 11 & 2 & 0 & 0 & 13 & 99  & 99  & 41  & 13  & 252 \\
Java       & 3 & 4 & 3 & 1 & 11 & 29 & 0 & 0 & 0 & 29 & 128 & 87  & 18  & 1   & 234 \\
Go         & 3 & 5 & 3 & 1 & 12 & 26 & 8 & 1 & 0 & 35 & 177 & 187 & 53  & 47  & 464 \\
C++        & 2 & 4 & 4 & 1 & 11 & 23 & 14 & 9 & 1 & 47 & 274 & 334 & 120 & 108 & 836 \\
JavaScript & 4 & 3 & 1 & 3 & 11 & 14 & 7 & 4 & 1 & 26 & 216 & 211 & 56  & 55  & 538 \\
\midrule
\textbf{Total} & \textbf{20} & \textbf{18} & \textbf{14} & \textbf{6} & \textbf{58}
& \textbf{103} & \textbf{31} & \textbf{14} & \textbf{2} & \textbf{150}
& \textbf{894} & \textbf{918} & \textbf{288} & \textbf{224} & \textbf{2,324} \\
\bottomrule
\end{tabular*}
\begin{tablenotes}[flushleft]
\item For L0/L1 and L2, S, M, L, and XL denote the generated code is $<500$, 500--1,999, 2,000--6,000, and $>6,000$ LOC, respectively.
\item For L3, the corresponding ranges are $<15$, 15--29, 30--49, and $\geq 50$ LOC.
\end{tablenotes}
\end{threeparttable}
\end{table*}

\textbf{Input Construction.}
For each task, we first extract task-relevant evidence from the Oracle repository. The extracted evidence depends on the task level and may include source code, test code, repository structure, public interfaces, target-file skeletons, function signatures, input-output information, and related repository entities. 
Using a fixed generation template, we then provide this structured evidence to Claude Sonnet 4.6 and instruct it to summarize the target behavior without introducing functionality unsupported by the evidence. The generated specification follows a predefined JSON schema and is validated for required-field completeness, target-behavior coverage, and consistency with extracted interfaces. The final input combines this semantic specification with the structural information permitted at that task level.

\begin{itemize}[leftmargin=10pt]
    \item \textbf{For L0 and L1}, we provide the complete Oracle repository and its tests to the specification model and ask it to summarize the repository objective, externally observable behavior, public entry points, major functional requirements, and engineering constraints. L0 uses the resulting repository specification alone, while L1 additionally includes the preserved repository skeleton.
    \item \textbf{For L2}, scripts first extract the hollowed target-file skeleton and the signatures of the functions to be generated. These artifacts, together with the complete Oracle implementation of the target file and relevant repository context, are used to generate a file-level description and per-function responsibility descriptions.
    \item \textbf{For L3}, scripts extract the target function signature, input-output information, and related interface information. These artifacts, together with the Oracle function implementation and local source context, are used to generate a behavior-oriented function description.
\end{itemize}
The Oracle implementation serves both as the semantic basis for constructing task specifications and as the factual reference for input-quality validation. During validation, each generated specification is compared with the corresponding Oracle implementation, tests, and context to assess its accuracy and sufficiency. The Oracle implementation is not exposed to the model performing the code-generation task; during evaluation, the generated artifact is integrated into an Oracle-derived repository environment and assessed through build and test execution rather than textual agreement with the Oracle code. Figure~\ref{fig:prompt-structure-examples} shows the example input for each task.

\textbf{Input Quality Assessment.}
We apply the same validation protocol across all task levels. 
Each constructed input is independently reviewed by GPT-5.4, Claude Sonnet 4.6, and DeepSeek-V4 Pro against the corresponding Oracle evidence.
Following established requirements-quality criteria~\cite{ref_iso29148,ref_requirements_quality_mapping}, we assess four dimensions: \textbf{completeness}, whether the input covers the information needed to implement the target artifact; \textbf{clarity}, whether the specification contains ambiguity or missing information; \textbf{consistency}, whether it faithfully reflects the Oracle implementation; and \textbf{implementability}, whether the provided behavioral and interface information is sufficient for reconstruction. 

For L0 and L1, the reviewer is given the repository specification together with the Oracle repository, relevant tests, and task metadata. For L2, the reviewer additionally examines the file-level and per-function descriptions against the target-file skeleton and complete Oracle file. For L3, the function description is checked against the target function and its surrounding source context.
The scores from the three reviewer models are aggregated into an overall quality score for each input. 
Inputs whose overall score does not exceed 3.5 are manually checked by three authors with at least three-year development experience against the reviewer feedback and Oracle evidence, revised to correct omissions, inaccuracies, or ambiguities, and then reassessed using the same procedure.
This process continues until every input included in the final benchmark achieves an overall quality score above 3.5. 
We have released all detailed prompts and results at our project homepage.

\subsection{Execution-Based Validation}
\label{sec:test-enhancement}
\benchmark{} adopts a unified execution-based validation framework across all task levels. To support consistent multilingual evaluation, language-specific processing is abstracted into four standardized interfaces: parser, slicer, sandbox execution, and static analysis.
The current release supports Python (pytest), Java (JUnit5), Go (\texttt{go test}), C++ (Google Test), and JavaScript (Jest)~\cite{ref_pytest,ref_junit,ref_gotest,ref_googletest,ref_jest}.
All build, test, and coverage procedures are executed in isolated Docker environments using each repository's native workflow. 
On top of this infrastructure, we first audit and strengthen test coverage for improving the capability for evaluating candidates' correctness, then apply level-specific integration and execution protocols, and finally perform Oracle validation.

\textbf{Coverage Auditing and Test Supplementation.} Execution-based evaluation is only meaningful when the available tests sufficiently exercise the implementation under evaluation. We therefore audit line and branch coverages over the target production-code regions before finalizing benchmark tasks. For each repository, we execute its native test suite and collect coverage using language-specific tools: \texttt{pytest-cov} for Python, \texttt{JaCoCo} for Java, \texttt{go test -coverprofile} for Go, \texttt{lcov} for C++, and \texttt{Jest/nyc} for JavaScript. The resulting reports are normalized into a unified coverage view across languages.
The initial 58 Oracle repositories show substantial variation in test quality. Coverage auditing indicates that only 27 repositories initially exceed 85\% line coverage, with an overall mean of 79.7\%; Java is particularly under-covered, with a mean of only 31.3\%. This indicates that directly reusing the original test suites would leave some repositories insufficiently constrained on their core implementation paths. The issue is particularly important for L2 and L3, where weak coverage over the target files or functions would reduce the benchmark's ability to distinguish generation quality. We therefore place under-covered repositories into a test-case supplementation stage.

We therefore supplement tests for under-covered repositories using Claude Sonnet 4.6. The model receives the repository source code, existing tests, and coverage reports, and is instructed to target uncovered functions, branches, and boundary cases while following the repository’s native testing framework. Generated tests are retained only after successfully passing the original test workflow, after which coverage is recomputed to verify the improvement. After supplementation, all 58 repositories exceed 85\% line coverage, and the overall mean increases from 79.7\% to 93.9\%.

\begin{figure}[t]
\centering
\includegraphics[width=1\linewidth]{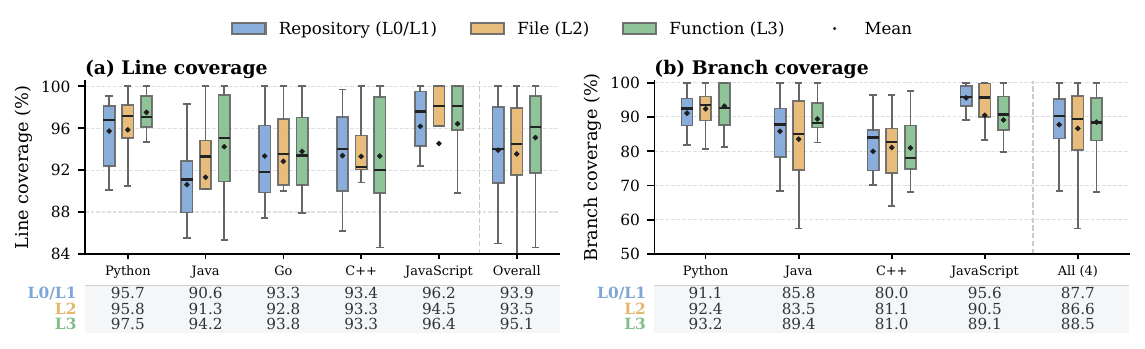}
\caption{Final line and branch coverage by language and level. Diamonds and aligned values denote mean coverage; Branch coverage for Go is unavailable due to tool limitations.}
\Description{Two side-by-side box plots show final line- and branch-coverage distributions for repository-, file-, and function-level evaluation scopes. Diamonds denote arithmetic means, and three aligned rows beneath each panel report the exact means by language and level. Go branch coverage is unavailable, so the overall branch group aggregates Python, Java, C++, and JavaScript.}
\label{fig:test-coverage-boxplot}
\end{figure}

Figure~\ref{fig:test-coverage-boxplot} reports coverage at the scope of each generation level: L0/L1 use whole-repository coverage, L2 restricts coverage to the target file, and L3 restricts coverage to the target function body. For L2/L3, task-level percentages are first averaged within each repository, so each repository contributes one observation to its language group and to the overall group. Panels (a) and (b) separately show the resulting line- and branch-coverage distributions.
Tasks or repositories without measurable branches are excluded rather than treated as zero. Go is marked unavailable because its standard coverage profile does not report branches, and the overall branch distribution therefore aggregates the four remaining languages.

\textbf{Level-specific execution protocols.} \benchmark{} applies two integration paths according to the generation target. 
All four levels follow the same execution-based principle, but complete-repository generation and local completion require different integration procedures.

\begin{itemize}[leftmargin=10pt]
    \item \textbf{For L0 and L1}, the generated artifact is a complete repository. L0 starts from an empty workspace, whereas L1 starts from the provided repository skeleton. The generated repository is executed in an isolated environment using interface-level functional tests. These tests exercise externally observable behavior through public interfaces such as CLIs, APIs, function entry points, and outputs, without assuming the internal structure of the Oracle implementation. The goal is to determine whether the generated artifact constitutes a runnable repository that satisfies the specified functionality.
    \item \textbf{For L2 and L3}, the generated artifact is a local completion. L2 generates the complete target file, whereas L3 generates the target function body. 
    The generated artifact is backfilled into the corresponding Oracle-derived repository and evaluated using all the associated test cases that involve the targeted file or function. This protocol requires the generated code to satisfy not only local functional behavior but also repository-level interfaces, dependencies, type constraints, and module interactions.
\end{itemize}

\textbf{Oracle validation.} Finally, we validate every benchmark task using the corresponding Oracle implementation through the same execution path later applied to generated artifacts. 
For L0/L1, the original repository is evaluated using the interface-level functional tests. For L2/L3, the reference file or function is backfilled into the repository and evaluated through the same build-and-test workflow.
All 2,590 tasks pass Oracle validation under their corresponding protocols. 
This confirms that task construction, artifact integration, and execution workflows are mutually consistent before model evaluation, reducing the likelihood that subsequent failures are caused by benchmark-construction artifacts rather than model behavior.

\section{Case Study Design}

In order to evaluate the performance of the latest methods and recent LLMs for code generation, we conducted a case study over our newly constructed benchmark.

\subsection{Research Questions}
\label{sec:questions}

\begin{itemize}[leftmargin=10pt]
    \item \textbf{RQ1:} How effective are existing methods in repository-level code generation across languages?
    \item \textbf{RQ2:} How effective are existing methods for file-level program generation across languages?
    \item \textbf{RQ3:} How effective are existing methods for function-level program generation across languages?
    \item \textbf{RQ4:} To what extent do different contexts affect generation performance?
\end{itemize}

\subsection{Studied Methods}
\label{sec:methods}

We evaluate three categories of code-generation systems: (1) \textbf{Direct generation models} produce code directly from a prompt or task specification without repository-level processing. (2) \textbf{General-purpose coding agents} can inspect and edit repositories autonomously but are not scope-specific, tailored to a specific generation scope. (3) \textbf{Specialized research methods} target a particular generation scope and embody explicit code-generation workflows or repository-context strategies. Including all three lets us compare specialized methods against one another, direct models, and general agents,
thereby also capturing recent advances in general-purpose large language models.

To select the specialized research methods systematically, we surveyed representative code-generation work published or released between 2023 and 2026. A candidate method was included only if it satisfied four criteria: (1) its generated artifact matches the target task level of our study; (2) it is technically distinct rather than a minor variant; (3) a stable and reproducible public implementation is available; and (4) adapting it to the five target languages does not require redesigning its core algorithm, keeping the evaluation feasible.
These criteria ensure task alignment, technical diversity, reproducibility, and cross-language feasibility. Table~\ref{tab:related-works} groups candidates by final-output scope, with boldface marking selected methods; details follow below.

\begin{table*}[!t]
\caption{Representative repository-aware code-generation methods considered in our study.}
\label{tab:related-works}
\centering
\resizebox{\columnwidth}{!}{
\begin{tabular}{lccllc}
\toprule
Method & Year & Venue & Scope & Technical path & Eval. lang. \\
\midrule
\multicolumn{6}{@{}l}{\textit{Function-oriented outputs}} \\
\textbf{RepoCoder} \cite{ref_repocoder} & 2023 & EMNLP & Function & Iterative retrieve-generate with draft query rewriting & Python \\
MonitorGuided \cite{ref_monitor_guided} & 2023 & NeurIPS & Function & Static-analysis monitors during decoding & Java, C\#, Rust \\
Repoformer \cite{ref_repoformer} & 2024 & ICML & Function & Selective RAG with retrieval gating & Python, Java, C\#, TS \\
DraCo \cite{ref_draco} & 2024 & ACL & Function & Repository data-flow graph context & Python \\
GraphCoder \cite{ref_graphcoder} & 2024 & ASE & Function & Coarse-to-fine CCG retrieval & Python, Java \\
CatCoder \cite{ref_catcoder} & 2025 & TOSEM & Function & Retrieval plus type-dependency prompting & Java, Rust \\
RLCoder \cite{ref_rlcoder} & 2025 & ICSE & Function & RL retriever with perplexity reward & Python, Java \\
\textbf{HCP} \cite{ref_hcp} & 2025 & AAAI & Function & Hierarchical context pruning & Python \\
ProjAPI \cite{ref_projapi} & 2025 & TSE & Function & Internal API inference for retrieval & Python, Java \\
\textbf{AlignCoder} \cite{ref_aligncoder} & 2025 & ASE & Function & Query enhancement plus AlignRetriever & Python, Java \\
CodexGraph \cite{ref_codexgraph} & 2025 & NAACL & Both & Graph-database repository indexing & Python, Java \\
\midrule
\multicolumn{6}{@{}l}{\textit{Repository-oriented outputs}} \\
MetaGPT \cite{ref_metagpt} & 2024 & ICLR & Repository & SOP-based multi-agent workflow & Python \\
\textbf{ChatDev} \cite{ref_chatdev} & 2024 & ACL & Repository & Chat-chain coordination of design, coding, and testing & Python, JS, etc. \\
CodePlan \cite{ref_codeplan} & 2024 & FSE & Repository & Dependency analysis with adaptive planning & C\#, Python \\
AgileCoder \cite{ref_agilecoder} & 2025 & Forge & Repository & Sprint-style iteration with dynamic code graph & Python \\
MegaAgent \cite{ref_megaagent} & 2025 & ACL & Repository & Dynamic agents with parallel decomposition & Python \\
RPG \cite{ref_rpg} & 2026 & ICLR & Repository & Repository Planning Graph with ZeroRepo & Python \\
\bottomrule
\end{tabular}
}
\parbox{\columnwidth}{\scriptsize\textit{Scope.} Function denotes localized completion within an individual method; Repository denotes complete-repository generation or repository-wide editing; Both denotes evaluation at both scopes. Boldface indicates methods selected for our study.}
\end{table*}

\begin{itemize}[leftmargin=10pt]
    \item \textbf{For L0 and L1:} 
    Both require a complete executable repository, differing only in whether a skeleton is supplied: L0 creates repositories from requirements, while L1 completes an existing skeleton while preserving its structure and interfaces. Both demand multi-file generation, dependency/build management, and cross-file consistency. We therefore evaluate general agentic methods \textbf{Codex} (GPT-5.4) and \textbf{Claude Code} (Claude Sonnet 4.6) at both levels. For L0 only, we add specialized \textbf{ChatDev}~\cite{ref_chatdev} with the same two LLM backends, since its collaboration mechanism adapts to our five languages without redesign; we exclude it from L1 because supplying a skeleton alters its core interface. Other baselines in Table~\ref{tab:related-works} are excluded due to heavy adaptation overhead, e.g., RPG would require reimplementing Python-centric parsing, skeleton construction, and test infrastructure for other languages.

    \item \textbf{For L2}:
    L2 requires jointly completing all non-trivial functions in one target file from a structured prompt containing the file description, function responsibilities, and target skeleton. We compare \textit{Agent}, where \textbf{Codex} (GPT-5.4) and \textbf{Claude Code} (Claude Sonnet 4.6) inspect the repository and edit the target file, with \textit{Direct LLM}, where the same models (\textbf{GPT-5.4} and \textbf{Claude Sonnet 4.6}) generate the whole file from the prompt alone. This isolates the benefit of agent-mediated repository inspection under a fixed task specification. Function-oriented retrieval methods in Table~\ref{tab:related-works} are excluded because they retrieve context for a single target function rather than jointly generating and integrating all implementations in a file.

    \item \textbf{For L3}: L3 requires generating a single target function body from its available context, e.g., functionality description. We selected three recent representative speciallized approaches: \textbf{RepoCoder}, \textbf{AlignCoder}, and \textbf{HCP}, which respectively represent \textit{iterative retrieval}, \textit{intent-aligned}, and \textit{post-retrieval context pruning} methods, respectively. We also include state-of-the-art LLMs as general-purpose baselines. Each method is evaluated with \textbf{GPT-5.4}, \textbf{Claude Sonnet 4.6}, and \textbf{DeepSeek-V4 Pro}, so that generator capacity is held constant and performance differences can be attributed to context-processing strategies rather than to the underlying LLM.

\end{itemize}

\subsection{Experimental Setup}

\subsubsection{Procedure}
For each task at a given level, we generate candidate outputs using the corresponding studied methods. Available information depends on task level and setting. Direct LLM methods receive only the structured prompt, agentic frameworks may inspect the exposed workspace and create or modify the target artifact within the prescribed scope, while specialized methods use their default code-generation settings unchanged for fair comparison. In particular, we allow automatic retries only for infrastructure failures, such as connection interruptions and request timeouts.

Generation and evaluation are separated. After generation, we freeze the final output and reintegrate it into an executable environment via the Section~\ref{sec:test-enhancement} artifact-integration procedure: L0/L1 outputs reconstruct complete repositories, L2 outputs replace target files, and L3 outputs are backfilled into target functions. All outputs are evaluated with the same dependencies, build commands, and test suite as the task. For L2/L3, evaluation is restricted to the associated test cases, while L0/L1 run the complete suite.

\subsubsection{Evaluation Metrics}
\label{sec:matrics}

We adopt three execution-based metrics commonly used in repository-level code-generation and software-engineering benchmarks~\cite{ref_swebench,ref_deveval,ref_projecteval}: (1) \textbf{Compilation Rate (\textit{Build})}, which indicates the proportion of all tasks whose generated artifacts compile successfully; (2) \textbf{Test Pass Rate (\textit{Test})}, which is conditional on successful build, the average fraction of associated test cases passed per task (build failures are excluded, so tasks with more tests are not weighted more); (3) \textbf{Full Pass Rate (\textit{Full})}, which is the proportion of all tasks that both compile and pass all associated tests, the strictest metric.

In particular, all studied methods are evaluated on the same set of tasks. Their outcomes are paired by task identifier, and statistical tests are conducted on task-level differences. Specifically, binary metrics \textit{Build} and \textit{Full} outcomes are analyzed using two-sided exact McNemar tests~\cite{ref_mcnemar}. Task-level \textit{Test} ratios use two-sided Wilcoxon signed-rank tests~\cite{ref_wilcoxon} over tasks that build successfully under both conditions.
When multiple related hypotheses are tested together, we apply Holm correction~\cite{ref_holm} with a significance level of $\alpha=0.05$. The corresponding statistical significance results are reported in the result analysis.

\subsubsection{Configurations and Environments}
\label{sec:config}

For specialized methods (RepoCoder, AlignCoder, HCP, ChatDev), we used their default settings with different LLM backends. All LLMs were accessed via online APIs. Codex and Claude Code ran through native CLIs with default reasoning and tool-use, and their internal sampling was not forced to match LLM-direct generation. For direct API calls and the final generation stage of specialized methods, we set temperature to 0.2, following HumanEval’s optimal pass@1 setting~\cite{ref_humaneval}. All other configurations followed the defaults of the corresponding methods or LLM interfaces. All detailed configurations are released in our project homepage.

All generation tasks run in mutually isolated workspaces. Evaluation is performed in Docker containers provided by our benchmark, with fixed dependencies, build configurations, and test commands for each repository. Outputs for the same task used identical integration procedures and execution environments, preventing environmental confounds. All experiments were conducted on a local machine equipped with an Intel Core i9 CPU, an NVIDIA GeForce RTX 4070 Laptop GPU, and 16 GB RAM, running Ubuntu 22.04.

\section{Evaluation Results}

\subsection{RQ1: Effectiveness in Repository-Level Code Generation}

Table~\ref{tab:l0l1-results} presents a comprehensive evaluation results of repository-level execution outcomes and resource usage at two levels of code generation tasks (L0 and L1). 

At L0, general-purpose coding agents generally outperform the specially designed ChatDev framework. Claude Code achieves the highest overall performance, with \textit{Build}, \textit{Test}, and \textit{Full} pass rates of 74.1\%, 91.0\%, and 22.4\%, respectively; Codex reaches 63.8\% \textit{Build} and 20.7\% \textit{Full}, and both general-purpose agents exceed 88\% \textit{Test}. In comparison, ChatDev-Sonnet and ChatDev-GPT achieve \textit{Build} rates of 56.9\% and 46.6\% and \textit{Full} rates of 17.2\% and 12.1\%, respectively. ChatDev consumes only 6.04M--7.42M tokens, fewer than Codex (10.8M) and Claude Code (74.27M), but still costs \$29.32--\$35.54, exceeding Codex-GPT at \$14.87. This may be because general-purpose agents obtain greater caching benefits when repeatedly analyzing similar contexts, whereas ChatDev has a lower cache hit rate and therefore remains costly. ChatDev also exhibits greater sensitivity to programming language: ChatDev-GPT reaches 92.3\% \textit{Build} on Python but remains below 50\% on each of the other four languages, raising concerns about its generality.

The choice of underlying LLM also significantly affects performance in both frameworks. Within ChatDev, the Sonnet backbone substantially outperforms the GPT backbone across all metrics, achieving 56.9\% \textit{Build}, 89.3\% Test, and 17.2\% Full pass rates, compared with 46.6\%, 84.8\%, and 12.1\% for GPT. Our manual inspection attributes this disparity to Sonnet’s more stable handling of local engineering constraints. Conversely, GPT struggles with dependency wiring and compatibility interfaces, such as Java incorrect imports and Go omitting required compatibility interfaces.
Across general agents, Claude Code generally achieves higher \textit{Build} and \textit{Full} pass rates than Codex. However, Codex is significantly more cost-effective, requiring substantially fewer tokens and costing less than half as much as Claude Code (\$14.87 vs. \$48.76).

As for L1, 
by providing the complete structure of the repository, both general agents perform better at L1 than at L0 regarding Full pass rate. Claude Code again achieves the highest Overall \textit{Build}, \textit{Test}, and \textit{Full} pass rates (62.1\%, 92.2\%, and 31.0\%, respectively) compared with Codex. Despite the high \textit{Test} pass rates, \textit{Full} pass rates remain around 30\% for both methods, indicating that full repository integration continues to pose significant challenges. Surprisingly, the \textit{Build} rate at L1 is substantially lower than at L0, indicating that general agents are better suited to generating programs freely than to working within a fixed structural implementation.
Furthermore, resource consumption increases drastically at L1: Claude Code’s token usage rises to 103.79M (costing \$102.53) and Codex’s to 26.71M (costing \$37.19). This suggests that L1-level repository generation, which takes the given repository structure as input, demands substantially greater computational resources.

\begin{summary}
    \textbf{RQ1 Summary:} (1) General-purpose coding agents outperform specially-designed ChatDev at the repository level: Claude Code achieves the strongest execution results, while Codex is more cost-efficient. (2) Repository skeletons improve full-pass but reduce build success and increase resource use, highlighting the need for both functional implementation and consistency across dependencies and structures.
\end{summary}

\begin{table*}[tbp]
\caption{L0/L1 repository-level code generation results and resource usage (58 tasks).}
\label{tab:l0l1-results}
\centering
\scriptsize
\def\msd#1#2{\ensuremath{#1^{\scriptscriptstyle\pm #2}}}

\def\bmsd#1#2{\ensuremath{\mathbf{#1}^{\scriptscriptstyle\mathbf{\pm #2}}}}

\renewcommand{\msd}[2]{#1}
\renewcommand{\bmsd}[2]{\textbf{#1}}
\resizebox{0.8\columnwidth}{!}{
\begin{threeparttable}
\begin{tabular}{lllcccccccc}
\toprule
\multirow{2}{*}{Level}& \multirow{2}{*}{Method}& \multirow{2}{*}{Metric}& \multicolumn{6}{c}{Execution (\%)} & \multicolumn{2}{c}{Resources} \\
\cmidrule(lr){4-9}\cmidrule(l){10-11}
 &  &  & C++ & Go & Java & JS & Python & Overall & Tokens (M) & Cost (\$) \\
\midrule
\multirow{12}{*}{L0}
& \multirow{3}{*}{{Claude Code (Sonnet)}}
  & \textit{Build} & \textbf{72.7} & \textbf{58.3} & \textbf{81.8} & \textbf{63.6} & \textbf{92.3} & \textbf{74.1} & \multirow{3}{*}{74.27} & \multirow{3}{*}{48.76} \\
& & \textit{Test}  & \msd{94.4}{4.5} & \msd{97.3}{2.3} & \msd{95.7}{5.4} & \msd{\textbf{91.7}}{10.8} & \msd{81.2}{14.4} & \bmsd{\textbf{91.0}}{11.3} & & \\
& & \textit{Full}  & \textbf{18.2} & 16.7 & \textbf{36.4} & \textbf{36.4} & 7.7 & \textbf{22.4} & & \\
\cline{2-11}
& \multirow{3}{*}{Codex (GPT)}
  & \textit{Build} & 54.5 & 50.0 & 72.7 & 45.5 & \textbf{92.3} & 63.8 & \multirow{3}{*}{10.80} & \multirow{3}{*}{14.87} \\
& & \textit{Test}  & \msd{\textbf{94.5}}{4.7} & \msd{\textbf{97.5}}{3.5} & \msd{95.9}{4.8} & \msd{87.1}{12.4} & \msd{76.5}{16.2} & \msd{88.5}{14.0} & & \\
& & \textit{Full}  & \textbf{18.2} & \textbf{33.3} & \textbf{36.4} & 9.1 & 7.7 & 20.7 & & \\
\cline{2-11}
& \multirow{3}{*}{{ChatDev (Sonnet)}}
  & \textit{Build} & 54.5 & 25.0 & 54.5 & \textbf{63.6} & 84.6 & 56.9 & \multirow{3}{*}{7.42} & \multirow{3}{*}{35.54} \\
& & \textit{Test}  & \msd{92.7}{8.8} & \msd{95.5}{3.7} & \msd{\textbf{96.5}}{4.2} & \msd{87.8}{16.0} & \msd{82.6}{14.2} & \msd{89.3}{13.0} & & \\
& & \textit{Full}  & \textbf{18.2} & 8.3 & 27.3 & \textbf{36.4} & 0.0 & 17.2 & & \\
\cline{2-11}
& \multirow{3}{*}{ChatDev (GPT)}
  & \textit{Build} & 27.3 & 33.3 & 27.3 & 45.5 & \textbf{92.3} & 46.6 & \multirow{3}{*}{6.04} & \multirow{3}{*}{29.32} \\
& & \textit{Test}  & \msd{87.1}{13.3} & \msd{96.9}{5.3} & \msd{95.0}{4.2} & \msd{55.4}{30.6} & \msd{\textbf{89.9}}{7.0} & \msd{84.8}{20.7} & & \\
& & \textit{Full}  & 9.1 & 25.0 & 9.1 & 0.0 & \textbf{15.4} & 12.1 & & \\
\midrule
\multirow{6}{*}{L1}
& \multirow{3}{*}{{Claude Code (Sonnet)}}
  & \textit{Build} & \textbf{45.5} & \textbf{75.0} & \textbf{27.3} & \textbf{54.5} & \textbf{100.0} & \textbf{62.1} & \multirow{3}{*}{103.79} & \multirow{3}{*}{102.53} \\
& & \textit{Test}  & \msd{\textbf{92.7}}{6.7} & \msd{97.6}{5.0} & \msd{\textbf{66.7}}{47.1} & \msd{\textbf{99.9}}{0.3} & \msd{\textbf{90.6}}{10.7} & \bmsd{92.2}{17.7} & & \\
& & \textit{Full}  & 0.0 & 58.3 & \textbf{18.2} & \textbf{45.5} & \textbf{30.8} & \textbf{31.0} & & \\
\cline{2-11}
& \multirow{3}{*}{Codex (GPT)}
  & \textit{Build} & \textbf{45.5} & \textbf{75.0} & \textbf{27.3} & \textbf{54.5} & 92.3 & 60.3 & \multirow{3}{*}{26.71} & \multirow{3}{*}{37.19} \\
& & \textit{Test}  & \msd{84.8}{17.1} & \msd{\textbf{98.2}}{2.8} & \msd{\textbf{66.7}}{47.1} & \msd{99.7}{0.8} & \msd{84.9}{12.6} & \msd{89.3}{19.5} & & \\
& & \textit{Full}  & 9.1 & 50.0 & \textbf{18.2} & \textbf{45.5} & 23.1 & 29.3 & & \\
\bottomrule
\end{tabular}
\begin{tablenotes}[flushleft]
\item Tokens and costs report LLM API usage, including cached tokens.
\end{tablenotes}
\end{threeparttable}
}
\end{table*}

\subsection{RQ2: Effectiveness in File-level Code Generation}

Table~\ref{tab:l2-results} presents the results for file-level code generation.
In this setting, 
we compare direct generation with Codex and Claude Code, which can inspect repository files and use tools before editing the target file. This comparison evaluates whether repository-level interaction helps a model produce files that integrate correctly with the surrounding codebase.

Codex with GPT achieves the highest overall \textit{Build} and \textit{Full} rates, at 88.0\% and 76.7\%, respectively. Compared with Direct using the same GPT backbone, Codex improves \textit{Build} by 17.3 percentage points and \textit{Full} by 21.4 percentage points. Its advantage in file-level build success is statistically significant under the exact McNemar test after Holm correction ($p=3.08\times10^{-5}$). The larger improvement in \textit{Full} than in \textit{Build} suggests that repository interaction contributes not only to resolving dependencies, but also to aligning the completed file with repository-specific behavior.

Claude Code with Sonnet, however, does not produce a clear overall improvement over Direct with the same backbone. Their \textit{Build} rates are 74.0\% and 73.3\%, respectively, while their \textit{Full} rates are 56.7\% and 57.3\%; the difference in build success is not statistically significant. To explain this result, we sampled and compared tasks on which Claude Code failed. The first recurring issue is execution instability. Among the 150 tasks, 20 do not produce evaluable outputs: some terminate abnormally, while others exceed the time limit (i.e., 10min). Because all methods are evaluated under a unified invocation protocol and execution budget, these outcomes indicate that Claude Code can become trapped in prolonged context exploration, repeated editing, or an execution process that fails to terminate on some complex tasks. In addition, some runs terminate normally while leaving methods unimplemented, showing that the agent does not always verify that file completion is finished before stopping. Furthermore, the failures also show that repository access can introduce new interface and consistency errors when the retrieved context is not used correctly, such as confusing parameter types, invoking undefined methods, and misinterpreting API return structures. Such mistakes lead to compilation failures or runtime exceptions. Repository access therefore supplies useful information but does not by itself ensure effective file completion.

All four configurations nevertheless achieve high overall \textit{Test} rates among build-successful tasks, ranging from 93.0\% to 97.4\%, whereas their \textit{Full} rates range from 55.3\% to 76.7\%. This gap shows that most build-successful files implement the main required behavior, but some still omit boundary cases, exceptional paths, or repository-specific constraints. Codex narrows this gap most effectively. Relative to Direct with GPT, it improves \textit{Full} in every language, including gains of 34.5 percentage points in Java and 23.0 points in Python. The Python result is particularly informative: Codex has a lower \textit{Build} rate than Direct (92.3\% \textit{vs} 100.0\%) but a substantially higher \textit{Full} rate (61.5\% \textit{vs} 38.5\%). The benefit of repository interaction is therefore not limited to making more files build; when the context is used effectively, it can also improve the behavioral correctness of compiled files.

The gains also vary across programming languages. Codex improves \textit{Build} most strongly for Java and C++, by 31.0 and 27.6 percentage points, respectively, whereas its gains are smaller for Go and JavaScript, whose direct-generation baselines already exceed 80\%. Claude Code improves \textit{Full} for Java, Go, and JavaScript, but degrades it for C++ and Python, with the largest decrease occurring on Python. These results demonstrate that file-level generation depends jointly on language conventions, repository structure, context-use strategy, and agent execution stability. Evaluation on a single programming language is therefore insufficient to characterize the effectiveness and robustness of repository-aware file generation.

\begin{summary}
    \textbf{RQ2 Summary:} (1) At the file level, Codex delivers the best overall performance and substantially improves build and full-pass rates over Direct generation with the same GPT backbone, underscoring the importance of repository interaction for correctly interpreting and preserving repository semantic constraints. (2) The large performance gap across languages highlights the need to evaluate the robustness of existing code generation methods.
\end{summary}

\begin{table*}[t]
\caption{L2 file-level code generation results and resource usage (150 tasks).}
\label{tab:l2-results}
\centering
\scriptsize
\def\msd#1#2{\ensuremath{#1^{\scriptscriptstyle\pm #2}}}
\def\bmsd#1#2{\ensuremath{\mathbf{#1}^{\scriptscriptstyle\mathbf{\pm #2}}}}
\renewcommand{\msd}[2]{#1}
\renewcommand{\bmsd}[2]{\textbf{#1}}
\resizebox{0.8\textwidth}{!}{%
\begin{threeparttable}
\begin{tabular}{llcccccccc}
\toprule
\multirow{2}{*}{Method}& \multirow{2}{*}{Metric}& \multicolumn{6}{c}{Execution (\%)} & \multicolumn{2}{c}{Resources} \\
\cmidrule(lr){3-8}\cmidrule(l){9-10}
 &  & C++ & Go & Java & JS & Python & Overall & Tokens (M) & Cost (\$) \\
\midrule
\multirow{3}{*}{Direct (GPT)} & \textit{Build} & 51.1 & 85.7 & 55.2 & 88.5 & \textbf{100.0} & 70.7 & \multirow{3}{*}{1.89} & \multirow{3}{*}{13.35} \\
 & \textit{Test} & \bmsd{99.5}{1.5} & \bmsd{99.6}{1.0} & \msd{99.8}{0.3} & \msd{96.1}{16.4} & \msd{87.4}{26.1} & \bmsd{97.4}{12.6} &  &  \\
 & \textit{Full} & 46.8 & 65.7 & 48.3 & 73.1 & 38.5 & 55.3 &  &  \\
 \cline{2-10}
\multirow{3}{*}{Direct (Sonnet)} & \textit{Build} & 63.8 & 80.0 & 65.5 & 80.8 & 92.3 & 73.3 & \multirow{3}{*}{3.14} & \multirow{3}{*}{16.72} \\
 & \textit{Test} & \msd{98.4}{4.3} & \msd{99.1}{3.5} & \bmsd{99.9}{0.2} & \msd{97.1}{11.5} & \msd{86.2}{26.8} & \msd{97.3}{11.3} &  &  \\
 & \textit{Full} & 53.2 & 62.9 & 62.1 & 65.4 & 30.8 & 57.3 &  &  \\
\cline{2-10}
\multirow{3}{*}{Codex (GPT)} & \textit{Build} & \textbf{78.7} & \textbf{94.3} & \textbf{86.2} & \textbf{96.2} & 92.3 & \textbf{88.0} & \multirow{3}{*}{23.60} & \multirow{3}{*}{28.55} \\
 & \textit{Test} & \msd{91.3}{26.6} & \bmsd{99.6}{1.4} & \bmsd{99.9}{0.2} & \bmsd{99.8}{0.6} & \bmsd{96.4}{5.9} & \msd{97.1}{14.7} &  &  \\
 & \textit{Full} & \textbf{66.0} & \textbf{82.9} & \textbf{82.8} & \textbf{88.5} & \textbf{61.5} & \textbf{76.7} &  &  \\
\cline{2-10}
\multirow{3}{*}{Claude Code (Sonnet)} & \textit{Build} & 57.4 & 85.7 & 75.9 & 84.6 & 76.9 & 74.0 & \multirow{3}{*}{63.28} & \multirow{3}{*}{61.12} \\
 & \textit{Test} & \msd{92.4}{24.0} & \msd{99.1}{2.3} & \msd{95.4}{14.5} & \msd{99.7}{0.7} & \msd{55.9}{46.0} & \msd{93.0}{22.8} &  &  \\
 & \textit{Full} & 46.8 & 65.7 & 69.0 & 69.2 & 15.4 & 56.7 &  &  \\
\bottomrule
\end{tabular}%
\begin{tablenotes}[flushleft]
\item Tokens and costs report LLM API usage, including cached tokens.
\end{tablenotes}
\end{threeparttable}
}
\end{table*}

\subsection{RQ3: Effectiveness in Function-level Code Generation}

\begin{table*}[t]
\caption{L3 function-level code generation results and resource usage (2,324 tasks).}
\label{tab:l3-full}
\centering
\def\msd#1#2{\ensuremath{#1^{\scriptscriptstyle\pm #2}}}
\def\bmsd#1#2{\ensuremath{\mathbf{#1}^{\scriptscriptstyle\mathbf{\pm #2}}}}
\renewcommand{\msd}[2]{#1}
\renewcommand{\bmsd}[2]{\textbf{#1}}
\resizebox{\textwidth}{!}{
\begin{threeparttable}
\begin{tabular}{ll ccc ccc ccc ccc}
\toprule
\multirow{2}{*}{Language} & \multirow{2}{*}{Metric} &
  \multicolumn{3}{c}{LLM Direct} &
  \multicolumn{3}{c}{RepoCoder} &
  \multicolumn{3}{c}{AlignCoder} &
  \multicolumn{3}{c}{HCP} \\
\cmidrule(lr){3-5}\cmidrule(lr){6-8}\cmidrule(lr){9-11}\cmidrule(lr){12-14}
& & GPT & Sonnet & DeepSeek & GPT & Sonnet & DeepSeek & GPT & Sonnet & DeepSeek & GPT & Sonnet & DeepSeek \\
\midrule
\multirow{3}{*}{Python} & \textit{Build} & 71.4 & 54.0 & 67.9 & 75.0 & \textbf{87.3} & 55.6 & 77.4 & 48.8 & 74.2 & 81.3 & 64.3 & 81.3 \\
 & \textit{Test} & \msd{84.1}{35.5} & \msd{77.5}{41.3} & \msd{67.9}{45.8} & \bmsd{85.3}{34.2} & \msd{71.2}{43.9} & \msd{77.4}{41.3} & \msd{58.2}{47.5} & \msd{78.2}{40.6} & \msd{66.0}{45.7} & \msd{82.6}{36.4} & \msd{83.7}{35.9} & \msd{76.2}{41.7} \\
 & \textit{Full} & 42.5 & 26.2 & 28.2 & 42.1 & 35.7 & 32.5 & 19.8 & 26.6 & 24.6 & \textbf{45.2} & 38.5 & 38.5 \\
\midrule
\multirow{3}{*}{Java} & \textit{Build} & 45.3 & 43.2 & 54.3 & 76.5 & 72.2 & \textbf{79.5} & 48.7 & 50.9 & 47.4 & 67.9 & 4.7 & 62.8 \\
 & \textit{Test} & \msd{99.1}{2.1} & \msd{99.6}{1.0} & \msd{98.6}{4.5} & \msd{99.6}{1.8} & \msd{98.7}{5.8} & \msd{99.4}{3.2} & \msd{99.5}{1.3} & \msd{99.6}{1.1} & \msd{98.9}{3.6} & \msd{99.6}{1.3} & \bmsd{99.7}{0.5} & \bmsd{99.7}{1.5} \\
 & \textit{Full} & 33.3 & 33.3 & 35.0 & \textbf{65.4} & 53.4 & 64.5 & 33.3 & 39.3 & 33.3 & 54.7 & 3.8 & 50.4 \\
\midrule
\multirow{3}{*}{Go} & \textit{Build} & 62.7 & 68.5 & 81.5 & \textbf{95.5} & 84.1 & 85.3 & 68.3$^\dagger$ & 59.9$^\dagger$ & 65.9$^\dagger$ & 87.3 & 58.2 & 61.9 \\
 & \textit{Test} & \msd{74.8}{43.3} & \msd{78.9}{40.8} & \msd{65.3}{47.6} & \msd{76.0}{42.7} & \msd{65.3}{47.6} & \msd{71.7}{45.0} & \msd{67.8}{46.7}$^\dagger$ & \bmsd{80.5}{39.5}$^\dagger$ & \msd{68.9}{46.3}$^\dagger$ & \msd{76.7}{42.2} & \msd{77.7}{41.6} & \msd{76.2}{42.5} \\
 & \textit{Full} & 46.6 & 53.7 & 52.4 & \textbf{72.2} & 54.5 & 61.0 & 45.9$^\dagger$ & 47.6$^\dagger$ & 44.8$^\dagger$ & 66.6 & 44.8 & 46.8 \\
\midrule
\multirow{3}{*}{C++} & \textit{Build} & 56.0 & 29.8 & 55.9 & \textbf{81.7} & 67.3 & 78.5 & 57.8$^\dagger$ & 74.5$^\dagger$ & 57.8$^\dagger$ & 79.4 & 54.3 & 66.7 \\
 & \textit{Test} & \msd{99.2}{4.5} & \msd{99.5}{3.6} & \msd{99.1}{5.1} & \msd{99.5}{4.2} & \msd{99.3}{4.2} & \msd{99.4}{4.7} & \msd{99.3}{4.3}$^\dagger$ & \msd{99.4}{5.4}$^\dagger$ & \msd{99.3}{3.4}$^\dagger$ & \msd{99.5}{3.6} & \msd{99.3}{5.2} & \bmsd{99.6}{1.9} \\
 & \textit{Full} & 52.4 & 29.1 & 52.2 & \textbf{78.5} & 62.9 & 75.4 & 54.9$^\dagger$ & 72.2$^\dagger$ & 53.9$^\dagger$ & 76.3 & 52.3 & 63.5 \\
\midrule
\multirow{3}{*}{JS} & \textit{Build} & 79.9 & 87.0 & \textbf{98.3} & 91.1 & \textbf{98.3} & 90.0 & 88.3$^\dagger$ & 86.6$^\dagger$ & 16.7$^\dagger$ & 89.0 & 58.6 & 77.5 \\
 & \textit{Test} & \msd{93.9}{17.8} & \msd{93.1}{19.4} & \msd{93.4}{19.2} & \msd{97.8}{11.3} & \msd{95.1}{15.8} & \msd{97.9}{8.8} & \msd{97.2}{11.8}$^\dagger$ & \bmsd{98.8}{7.2}$^\dagger$ & \msd{97.3}{10.4}$^\dagger$ & \msd{98.3}{9.4} & \msd{97.5}{8.5} & \msd{97.3}{11.5} \\
 & \textit{Full} & 55.0 & 56.3 & 64.3 & \textbf{77.3} & 69.0 & 71.6 & 68.6$^\dagger$ & 73.8$^\dagger$ & 10.2$^\dagger$ & 76.6 & 47.2 & 62.8 \\
\midrule
\multirow{3}{*}{\textbf{Overall}} & \textit{Build} & 63.5 & 54.7 & 71.9 & \textbf{85.4} & 80.5 & 80.1 & 68.2 & 69.2 & 50.6 & 82.3 & 52.2 & 69.4 \\
 & \textit{Test} & \msd{91.0}{26.6} & \msd{89.6}{28.6} & \msd{86.4}{32.5} & \msd{92.5}{25.5} & \msd{87.6}{31.3} & \msd{91.5}{27.0} & \msd{87.3}{32.0} & \bmsd{94.3}{22.2} & \msd{85.9}{33.8} & \msd{92.6}{25.3} & \msd{91.9}{25.9} & \msd{91.9}{26.2} \\
 & \textit{Full} & 48.8 & 40.4 & 50.7 & \textbf{71.7} & 58.7 & 65.9 & 50.3 & 59.4 & 36.7 & 68.9 & 43.2 & 56.0 \\
\cmidrule(l){2-14}
  & Tokens (M) & 17.57 & 25.11 & 17.33 & 182.53 & 237.54 & 206.03 & 62.13 & 98.23 & 71.94 & 97.09 & 125.04 & 99.79 \\
  & Cost (\$) & 46.00 & 82.19 & 28.05 & 370.22 & 728.37 & 614.12 & 118.43 & 363.87 & 65.35 & 165.40 & 399.35 & 124.46 \\
\bottomrule
\end{tabular}
\begin{tablenotes}
\item $\dagger$: cross-language adaptation of AlignRetriever beyond its original Python/Java training languages.
\item Tokens and costs report LLM API usage, including cached tokens.
\end{tablenotes}
\end{threeparttable}
}
\end{table*}

Function-level code generation is one of the most widely studied settings in code generation. We evaluate three representative repository-aware methods, RepoCoder, AlignCoder, and HCP, together with LLM direct generation on \benchmark{}. Each method is paired with GPT-5.4, Claude Sonnet 4.6, and DeepSeek-V4 Pro, enabling comparisons across both context-processing methods and generator backbones. Table~\ref{tab:l3-full} reports the results over 2,324 L3 tasks.

Repository context provides measurable benefits, although the magnitude of the improvement depends on how the retrieved information is processed. Under the same GPT backbone, RepoCoder and HCP significantly outperform Direct on \textit{Build}, \textit{Test}, and \textit{Full} (all Holm-adjusted $p$<3.3$\times$$10^{-7}$). AlignCoder provides a smaller improvement and achieves statistical significance only on \textit{Build} ($p$=6.0$\times$$10^{-4}$). These results show that repository context can improve function generation, but simply introducing a retrieval pipeline does not guarantee consistent gains across all metrics.

RepoCoder is particularly notable because it is the earliest and structurally simplest repository-aware method evaluated in this study, yet RepoCoder with GPT achieves the highest overall \textit{Build} and \textit{Full} rates, at 85.4\% and 71.7\%, respectively. It significantly outperforms AlignCoder with GPT on all three metrics (all Holm-adjusted $p$<4.3$\times$$10^{-15}$) and outperforms HCP with GPT on \textit{Build} and \textit{Full} ($p$=1.8$\times$$10^{-3}$ and $p=$1.59$\times$$10^{-2}$, respectively). The effectiveness of a method is therefore not determined by its recency or pipeline complexity. 
A simpler, earlier retrieval method still performs best with GPT, suggesting that improving the model's ability to understand, select, and use repository context may yield greater gains than continually adding retrieval, alignment, and filtering stages. Retrieval remains valuable, but its benefit depends on the underlying model's capacity to exploit the supplied context.

Table~\ref{tab:l3-full} also reveals a persistent gap between test-pass rate and full-pass rate. \textit{Test} is the average fraction of tests passed among build-successful tasks, whereas \textit{Full} is the proportion of all tasks that pass the complete test suite. Across most method--model combinations, build-successful functions pass a large fraction of tests, while substantially fewer tasks satisfy the complete suite. This pattern indicates that many generated functions implement the main target behavior without achieving complete functional correctness.

Under direct generation, GPT, Sonnet, and DeepSeek obtain \textit{Test} rates of 91.0\%, 89.6\%, and 86.4\%, respectively, while their \textit{Full} rates are 48.8\%, 40.4\%, and 50.7\%. The same separation appears for repository-aware methods: RepoCoder with GPT and HCP with GPT reach \textit{Test} rates of 92.5\% and 92.6\%, but their \textit{Full} rates are 71.7\% and 68.9\%. These results suggest that LLMs generally cover the main execution paths but may not account fully for all task requirements. Generated code can behave correctly on routine inputs while failing to handle some boundary conditions, exceptional paths, or repository-specific behavioral constraints. Consequently, a small number of failed tests can prevent an otherwise largely correct implementation from satisfying the strict full-pass criterion. The gap between \textit{Test} and \textit{Full} therefore reflects the remaining distance between partial functional correctness and complete task correctness.

As at repository and file levels, performance at function level also varies substantially across languages, underscoring the need for unified multi-language execution protocols. Finally, the results demonstrate that repository-aware generation also consumes far more tokens: with GPT, Direct uses 17.57M, while AlignCoder, HCP, and RepoCoder use 62.13M, 97.09M, and 182.53M, which is about 3.5x, 5.5x, and 10.4x against Direct. To conclude, retrieval and multi-stage context processing can improve execution but add substantial overhead, so repository-aware methods should be assessed jointly on effectiveness, cross-language stability, and computational cost.

\begin{summary}
    \textbf{RQ3 Summary:} 
    (1) At the function level, repository context helps, but performance gains depend on the context-processing method and backbones. RepoCoder, though earliest and structurally simplest, leads build and full-pass rates with GPT. (2) The persistent gap between test and full-pass rates, token overhead, and cross-language variation show that repository-aware methods need evaluation beyond aggregate accuracy.
\end{summary}

\subsection{RQ4: The Impact of Context on Code Generation Performance}

RQ4 examines how context form affects code generation via two comparisons. First, L0 (complete repository from an empty workspace) \textit{versus} L1 (with a repository skeleton) isolates the effect of structural repository context. Second, L2-to-L3 \textit{versus} L3 Direct isolates related implementation context at the function level: at L2, the model generates the full target file using other functions, interfaces, and implementations as context; we then extract the target function and re-evaluate it with the same function-level procedure as L3 (we refer to this as L2-to-L3), whereas L3 Direct generates the same function without such context. We align the shared tasks pairwise to determine whether related implementation context improves function-level generation.

\begin{figure}[t]
  \centering
  \subfigure[L0 \textit{vs} L1: generate complete repositories.]{
     \includegraphics[width=0.48\columnwidth]{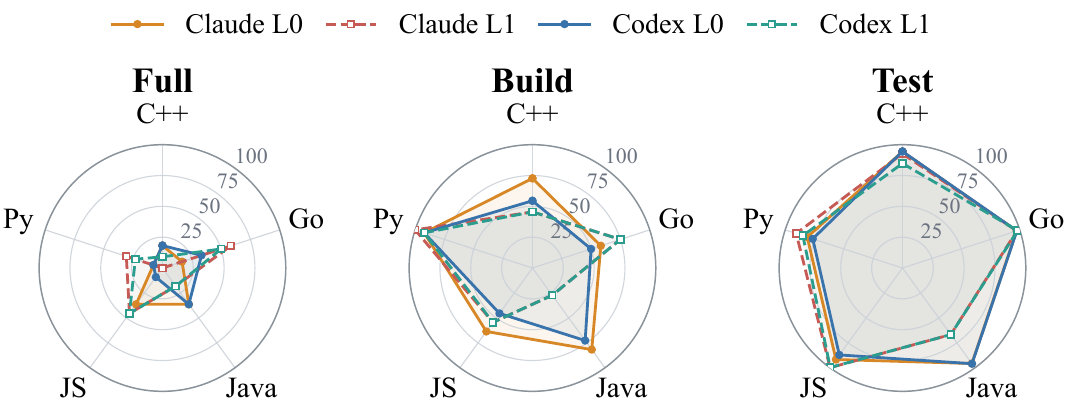}
     \label{fig:l0l1-language-radar}
  }\hfill
  \subfigure[L2-to-L3 \textit{vs} L3: generate individual functions.]{
     \includegraphics[width=0.48\columnwidth]{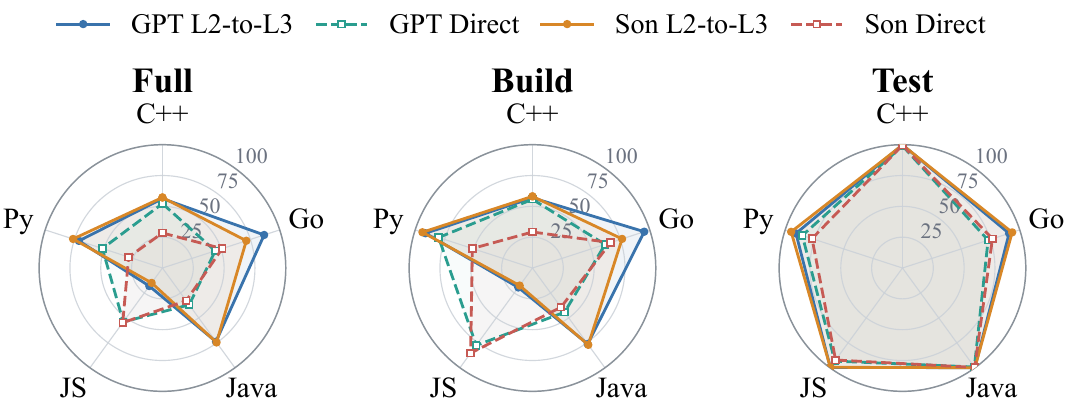}
     \label{fig:l2tol3-direct-radar}
  }
  \caption{Code generation comparison with different contexts across languages and metrics(\%)}
  \label{Fig.d4j}
\end{figure}

Figure~\ref{fig:l0l1-language-radar} compares Claude Code and Codex at L0 and L1. For both methods, providing a repository skeleton increases \textit{Full} and \textit{Test} but decreases \textit{Build}. From L0 to L1, the full-pass rate rises from 22.4\% to 31.0\% for Claude Code and from 20.7\% to 29.3\% for Codex, an increase of 8.6 percentage points in both cases. The test-pass rate also rises from 91.0\% to 92.2\% and from 88.5\% to 89.3\%, respectively. In contrast, the build-success rate decreases from 74.1\% to 62.1\% for Claude Code and from 63.8\% to 60.3\% for Codex. A repository skeleton exposes file organization, module boundaries, and interface definitions, helping the methods identify the functionality required from each component and allowing more repositories to pass the complete test suite. However, the generated code must also conform to a fixed directory structure, interfaces, and dependency relationships; inconsistencies with these constraints can prevent a repository from building. This explains why L1 improves the full-pass rate while reducing the build-success rate.

The second comparison further examines the effect of related implementation context on function generation. We align the two result sets by task identifier, yielding 1,891 shared tasks.
Figure~\ref{fig:l2tol3-direct-radar} shows that related implementation context increases the full-pass rate for both models. For GPT-5.4, the rate rises from 50.1\% to 55.0\%, a statistically significant increase of 4.9 percentage points ($p$=5.31$\times$$10^{-4}$). For Claude Sonnet 4.6, it rises from 40.2\% to 51.6\%, a statistically significant increase of 11.4 percentage points ($p$=2.24$\times$$10^{-14}$). These results indicate that related implementations in the target file provide valuable information (e.g., interface usage and local behavior), increasing the likelihood that the generated function satisfies the complete test suite.

The results demonstrate that both context forms improve full-pass performance but act differently. Related implementation context supplies interface and behavioral information, significantly raising function-level full-pass rates for both models. A repository skeleton exposes file structures and interfaces, helping satisfy functional requirements while imposing structural constraints that can increase build failures. The value of context therefore depends not merely on quantity but on relevance and the system’s ability to follow conveyed constraints.

\begin{summary}
    \textbf{RQ4 Summary:} 
    Although both repository skeletons and related implementation context improve full-pass performance, the benefit of such contexts depends on its relevance to the target task and on the method's ability to use it correctly.
\end{summary}

\section{Threats to Validity}

We identified several potential threats to the validity of our study. First, regarding repository selection, our benchmark favors actively maintained open-source repositories with mature engineering practices and testing infrastructure. Consequently, our findings may not generalize directly to private software, legacy systems, or other types of repositories. Second, regarding benchmark quality, we use Claude Sonnet 4.6 to generate repository-, file-, and function-level task descriptions from the Oracle implementations and associated evidence. Because automatically generated descriptions may not fully capture the intended functionality, we further use three recent LLMs (\textit{ref.} Section~\ref{sec:improvement})
to independently review and score their quality. Low-scoring tasks are manually inspected and revised to reduce omissions, ambiguities, and inconsistencies. Finally, because the benchmark is derived from public open-source repositories that may appear in the pretraining corpora of the evaluated models, training-data contamination or memorization cannot be fully excluded. We reconstruct task descriptions, withhold target implementations, and execute all generated artifacts in controlled environments, but these precautions do not eliminate this risk. The reported results should therefore be interpreted as measurements of method performance on public, real-world repository tasks, rather than as contamination-free estimates of generalization.

\section{Conclusion}

In this paper, we introduced \benchmark{}, a \textit{multilingual}, \textit{multi-granularity} benchmark for code generation in real-world software repositories. It spans five programming languages and defines four task levels, ranging from whole-repository generation to function-level completion. A unified build-and-test protocol enables direct generation, repository-aware methods, and coding agents to be evaluated under consistent execution criteria. Using this benchmark, we conducted a large-scale empirical study of existing code-generation approaches across three major categories: LLM-direct generation, state-of-the-art specialized methods, and general-purpose agentic frameworks. The results reveal several key findings, including the unsatisfactory performance of existing methods, the importance of effective context utilization, and the trade-off between cost and effectiveness. They also underscore the need for an execution-driven, multilingual, and multi-granularity benchmark to support unified evaluation of code-generation methods.

\section{Data Availability}

We have released our benchmark, all experimental configurations (including used prompts and scripts, etc.) and evaluation results (including benchmark quality assessments, detailed statistical results, and other unreported detailed analysis) to facilitate the replication and further comparison for future research in this domain at \homepage{}.

\bibliographystyle{ACM-Reference-Format}
\bibliography{references}

\end{document}